\documentclass[12pt,tightenlines,nofootinbib]{revtex4}
\usepackage{amsfonts,amsmath,amssymb}
\usepackage{txfonts}
\usepackage{graphicx}
\usepackage{color}
\usepackage{wrapfig,lipsum,booktabs}
\usepackage{floatrow,float}
\usepackage{epstopdf}
\usepackage{hyperref}
\usepackage{stmaryrd}
\usepackage{mathrsfs,mathtools}
\usepackage{ulem}

\begin{document}
	
	\title{Learning Orientations: a Discrete Geometry Model}
	\author{Y. Dabaghian}
	\affiliation{Department of Neurology, The University of Texas McGovern Medical School, 6431 Fannin St, 
		Houston, TX 77030\\ $^{*}$e-mail: Yuri.A.Dabaghian@uth.tmc.edu}
	\vspace{17 mm}
	\date{\today}
	
\begin{abstract}
In the mammalian brain, many neuronal ensembles are involved in representing spatial structure of the 
environment. In particular, there exist cells that encode the animal's location and cells that encode
head direction. A number of studies have addressed properties of the spatial maps produced by these two 
populations of neurons, mainly by establishing correlations between their spiking parameters and geometric
characteristics of the animal's environments. The question remains however, how the brain may intrinsically
combine the direction and the location information into a unified spatial framework that enables animals'
orientation. Below we propose a model of such a framework, using ideas and constructs from algebraic topology
and synthetic affine geometry.
\end{abstract}
	
\maketitle
	
\section{Introduction and background}
\label{sec:intro}

\begin{figure}[h!]
	\centering
	\includegraphics[scale=0.82]{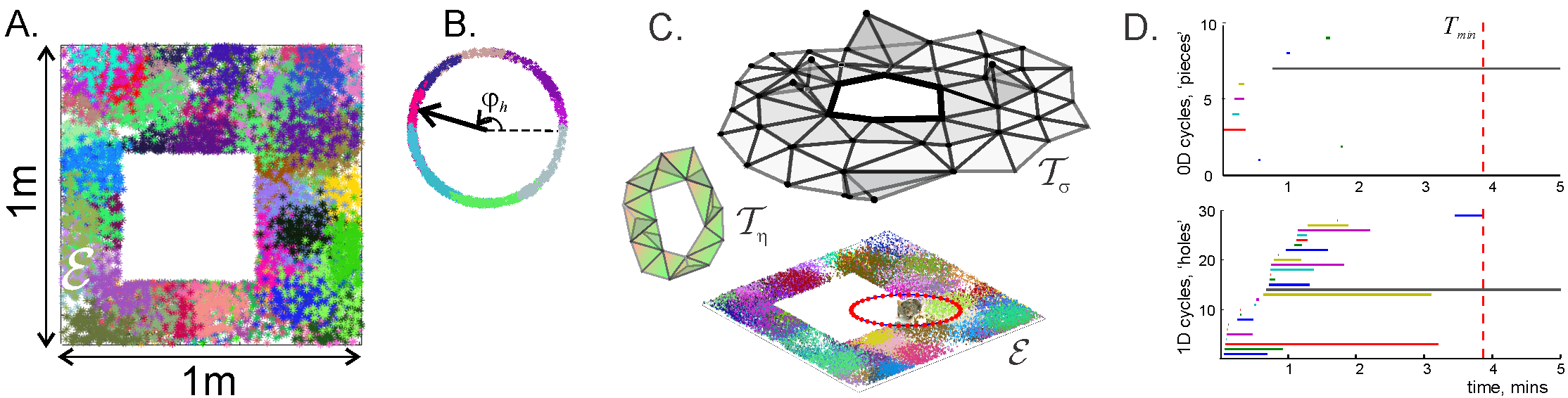}
	\caption{{\footnotesize\textbf{Basic topological constructions}. 
			\textbf{A}. Simulated place field map $M_\mathcal{E}$ with place fields scattered randomly in a 
			$1\times 1$ m square environment $\mathcal{E}$ with a square hole in the middle. Clusters of dots 
			of a particular color represent individual place fields. 
			\textbf{B}. A head direction field map $M_{S^1}$ covers the space of directions, $S^1$. Clusters
			of colored dots mark specific head direction fields $\upsilon_{h}$, centered each at its preferred
			angle $\varphi_h$. 
			\textbf{C}. The net pool of place cell coactivities is represented by the coactivity complex 
			$\mathcal{T}_{\sigma}(t)$ (top right), which provides a developing topological representation of
			the	environment $\mathcal{E}$ (bottom). The head direction cells map a circular space of directions
			$S^1$ (shown as a ring around the rat). The net pool of head direction cell activities is schematically
			represented the coactivity complex $\mathcal{T}_{\eta}(t)$.
			\textbf{D}. The timelines of the separate pieces (top panel) and holes (bottom panel) in the 
			complex $\mathcal{T}_{\sigma}(t)$ are shown as horizontal bars. At the onset of the navigation,
			$\mathcal{T}_{\sigma}(t)$ contains many spurious topological defects that disappear after a certain 
			``learning period" $T_{\min}^{\sigma}$, leaving behind a few persistent loops that define the 
			topological shape of $\mathcal{T}_{\sigma}(t)$ \cite{PLoS}. Similar behavior is exhibited by the
			head direction coactivity complex $\mathcal{T}_{\eta}(t)$.
		}}
	\label{fig:PFmap}
\end{figure}

\textbf{Spatial cognition} in mammals is based on an internalized representation of space---a \textit{cognitive
map} \footnote{Throughout the text, terminological definitions are given in \textit{italics}.} that emerges 
from neuronal activity in several regions of the brain \cite{OKeefe,Derdikman,Tolman}.
The type of information encoded by a specific neuronal population is discovered by establishing correspondences 
between its spiking parameters and spatial characteristics of the environment. For example, ascribing the 
$xy$-coordinates to every spike produced by the hippocampal principal neurons according to the animal's (in 
the experiments, typically rat's) position at the moment of spiking, produces distinct clusters, indicating 
that these neurons, the so-called \textit{place cells}, fire only within specific locations---their respective 
\textit{place fields} \cite{Dostrovsky,Best2}. 
The layout of the place fields in a spatial domain $\mathcal{E}$---the \textit{place field map} $M_{\mathcal{E}}$ 
(Fig.~\ref{fig:PFmap}A)---thus defines the temporal order of the place cells' spiking activity during the animal's
navigation, which is a key determinant of the cognitive map's structure. Hence, tagging the spikes with the location
information can be viewed as a mapping from a cognitive map $\mathcal{C}$ into the navigated space,
\begin{subequations}
	\label{map}
	\begin{align}
	\begin{split}
	f_{\sigma}: \mathcal{C}\to \mathcal{E},
	\label{mapPC}
	\end{split}
	\tag{1$\sigma$}
	\intertext{referred to as \textit{spatial mapping} in \cite{SchemaS}.
		Similarly, tagging the spikes produced by certain neurons in the postsubiculum (and in few other brain
		regions \cite{Taube,Wiener}) with the rat's head direction angle $\varphi$ produces clusters in the
		space of planar directions---the circle $S^1$, thus defining a mapping}
	\begin{split}
	f_{\eta}: \mathcal{C} \to S^{1}.
	\label{mapHD}
	\end{split}
	\tag{1$\eta$}
	\end{align}
\end{subequations}
The angular domains in which specific \textit{head direction cells} become active can be viewed as \textit{head
direction fields} in $S^1$, similar to the hippocampal place fields in the navigated space. The corresponding
\textit{head direction map}, $M_{S^1}$, determines the order in which the head direction cells spike during the
rat's movements (Fig.~\ref{fig:PFmap}B, \cite{TaubeGood,Muller}).

The preferred angular domains depend weakly, if at all, on the rat's position, just as place fields are overall
decoupled from the head or body orientation (see however \cite{Jercog,Rubin}). Thus, the following discussion
will be based on the assumption that both cell populations contribute to an \textit{allocentric} representation
of the ambient space: the place cells encode a topological map of locations \cite{Gothard1,Alvernhe1,Alvernhe2,Alvernhe3,
eLife,Wu}, whereas head direction cells augment it with angular information \cite{Taube2,Valerio,McNPth,Savelli}.

\textbf{Topological model}. The physiological and the computational mechanisms by which a cognitive map comes
into existence remain vague \cite{McNCog,Kropff,Viewpoints}. However, certain insights into its structure can
be obtained through geometric and topological constructions. For example, a place field map $M_{\mathcal{E}}$
can be viewed as a \textit{cover} of the navigated environment $\mathcal{E}$ by the place fields $\upsilon_i$,
\begin{equation}
\setcounter{equation}{2}
\mathcal{E}=\cup_i\upsilon_i,
\label{PFcover}
\end{equation} 
and used to link the topology of $\mathcal{E}$ to the topological structure of the cognitive map $\mathcal{C}$.
Indeed, according to the Alexandrov-\v{C}ech theorem, if every nonempty set of overlapping place fields,
$\upsilon_{i_0,i_1,\ldots,i_n}\equiv\upsilon_{i_0}\cap\upsilon_{i_1}\cap\ldots\cap\upsilon_{i_n}=\upsilon_{i_0,
	i_1,\ldots,i_n}\neq\varnothing$, is represented by an abstract simplex, 
$\nu_{i_0,i_1,\ldots,i_n}=[\upsilon_{i_0},\upsilon_{i_1},\ldots,\upsilon_{i_n}]$, then the homologies of the 
resulting simplicial complex $\mathcal{N}_{\sigma}$---the \textit{nerve} of the map $M_{\mathcal{E}}$---match 
the homologies of the underlying space $H_{\ast}(\mathcal{N}_{\sigma})=H_{\ast}(\mathcal{E})$, provided that
all the overlaps $\upsilon_{i_0,i_1,\ldots,i_n}$ are contractible. This implies that $\mathcal{N}_{\sigma}$ and
$\mathcal{E}$ have the same \textit{topological shape}---same number of connectivity components, holes, cavities,
tunnels, etc. \cite{Hatcher}. The same line of arguments allows relating the head direction map with the topology
of the space of directions $S^1$ (Fig.\ref{fig:PFmap}C).

It must be emphasized however, that reasoning in terms of place and head direction fields may not capture
the brain's intrinsic principles of processing spiking information, e.g., explain how either the location
or the direction signals contribute to animal's spatial awareness, because the experimentally constructed
firing fields are nothing but artificial constructions used to in interpret and visualize spiking data 
\cite{Hargreaves,Sargolini}. Addressing the brain's intrinsic space representation mechanisms requires carrying
the analyses directly in terms of spike times, without invoking auxiliary correlates between neuronal activity
and the observed environmental features.

Fortunately, the approach motivated by the nerve theorem can be easily transferred into a ``spiking" format.
Indeed, one can view a combination of the coactive place cells---a \textit{cell assembly} 
\cite{Hebb,Harris,Syntax}---as an abstract \textit{coactivity simplex},
\begin{equation}
\sigma_i=[c_{i_{0}},c_{i_1},\ldots,c_{i_n}]
\label{sigma}
\end{equation}
that activates when the rat crosses its \textit{simplex field} $\upsilon_{\sigma_i}$---a domain where all the
cells $c_i\in\sigma_i$ are coactive \cite{Curto}. By construction, this domain is defined by the overlap of the
corresponding place fields $\upsilon_{\sigma_i}=\upsilon_{i_0,i_1,\ldots,i_n}$, and may hence be viewed as the
the projection of $\sigma_i$s into $\mathcal{E}$ under the mapping (\ref{mapPC}). Note that if two coactivity
simplexes overlap, their respective fields also overlap, $\sigma_i \cap \sigma_j\neq\varnothing\Leftrightarrow
\upsilon_{\sigma_i}\cap\upsilon_{\sigma_j}\neq\varnothing$. Thus, if a cell $c_i$ is shared by a set $U_i$ of
simplexes, $U_i=\{\sigma:\sigma\cap c_i\neq\varnothing\}$, then its place field is formed by the union of the
corresponding $\sigma$-fields, $$\upsilon_i=\cup_{\sigma\in U_i}\upsilon_{\sigma}.$$

If a simplex $\sigma_i$ first appears at the moment $t_i$, then the net pool of neuronal activities produced by
the time $t$ gives rise to a time-developing simplicial \textit{coactivity complex} $$\mathcal{T}_{\sigma}(t)
=\cup_{t_i<t}\sigma_i$$ that inflates ($\mathcal{T}_{\sigma}(t)\subseteq\mathcal{T}_{\sigma}(t')$ for $t<t'$),
and eventually \textit{saturates}, converging to the nerve complex's structure, i.e., $\mathcal{T}_{\sigma}(t)
\approx \mathcal{N}_{\sigma}$, for $t\gtrsim T^{\sigma}_{\ast}$. Analyses based on simulating rat's moving
through randomly scattered place fields show that, e.g., for a small environment $\mathcal{E}$ illustrated on
Fig.~\ref{fig:PFmap}A, the rate of new simplexes' appearance slacks in about $T^{\sigma}_{\ast}\approx 6$ 
minutes \cite{SchemaS}, which provides an estimate for the time required to map $\mathcal{E}$.

The topological dynamics of $\mathcal{T}_{\sigma}(t)$ can be described using Persistent Homology theory
\cite{Zomorodian,Edelsbrunner,Kang}, which allows identifying the ongoing shape of $\mathcal{T}_{\sigma}(t)$
based on the times of its simplexes' first appearance. Typically, $\mathcal{T}_{\sigma}(t)$ starts off with
numerous topological defects that tend to disappear as the information provided by the spiking place cells
accumulates (see \cite{PLoS,Arai,Hoffman,Basso} and Fig.~\ref{fig:PFmap}D). Hence the minimal period 
$T_{\min}^{\sigma}$ required to recover the ``physical" homologies $H_{\ast}(\mathcal{E})$ provides an 
estimate for the time necessary to learn topological connectivity of the environment, which, for the case
illustrated on Fig.~\ref{fig:PFmap}A, is about $T_{\min}^{\sigma}\approx 4-5$ minutes \cite{PLoS,Arai,Basso,Hoffman,Alvernhe,Piet}.

Importantly, the coactivity complex may be used not only as a tool for estimating learning timescales, but 
also as a schematic representation of the cognitive map's developing structure, providing a context for 
interpreting the ongoing neuronal activity. Indeed, a consecutive sequence of $\sigma$-fields visited by the
rat,
\begin{equation}
\Upsilon=\{\upsilon_{i_0},\upsilon_{i_1},\ldots,\upsilon_{i_n},\ldots\},
\label{discrtraj}
\end{equation} 
captures the shape of the underlying physical trajectory $s\subset\Upsilon$ \cite{Guger,Jensen,Frank,Brown,ZhangRec}. 
The corresponding chain of the place cell assemblies ignited in the hippocampal network is represented by the 
\textit{simplicial path}
\begin{equation}
\tilde{\sigma} =\{\sigma_1,\sigma_2,\ldots,\sigma_n,\ldots\}.
\label{sigmapath}\tag{5$\sigma$}
\end{equation}
The fact that this information allows interpreting certain cognitive phenomena \cite{Pfeiffer1,Johnson,Dragoi}
suggests that the animal's movements are faithfully monitored by neuronal activity, i.e., that in sufficiently
well-developed complexes (e.g., for $t>T_{\min}^{\sigma}$) simplicial paths capture the shapes of the underlying
trajectories \cite{Guger,Jensen,Frank,Brown}. For the referencing convenience, this assumption is formulated as
two model requirements:

\vspace{2mm}
\textbf{R1. Actuality}. 
\textit{At any moment of time, there exists an active assembly $\sigma$ that represents the animal's
	current location.}

\vspace{2mm}

\textbf{R2. Specificity}.
\textit{Different place cell assemblies represent different domains in $\mathcal{E}$, i.e., $\sigma$-simplexes
	serve as unique indexes of the animal's location in a given map $\mathcal{C}$.}
\vspace{2mm}

An implication of these requirements is that the simplex fields cover the explored surfaces (\ref{PFcover}) and
that if the consecutive simplexes in (\ref{sigmapath}) are \textit{adjacent}, i.e., no simplexes ignite between
$\sigma_i$ and $\sigma_{i+1}$ (schematically denoted below as $\sigma_i\multimapdotboth\sigma_{i+1}$) then the
corresponding $\sigma$-fields are adjacent or overlap.

\vspace{2mm}
\textbf{Head orientation map}. Using the same line of arguments, one can deduce the topology of the space of 
directions by building a dynamic \textit{head direction coactivity complex} $\mathcal{T}_{\eta}(t)$ from the 
simplexes $$\eta_j=[h_{j_1},\ldots,h_{j_l}],$$ which designate the assemblies of head direction cells $h_{j_1}, 
h_{j_2},\ldots,h_{j_l}$. If a simplex $\eta_j$ first activates at the moment $t_j$, then $$\mathcal{T}_{\eta}(t)
=\cup_{t_{j}\leq t}\eta_j.$$ 
As the complex $\mathcal{T}_{\eta}(t)$ develops, it forms a stage for representing the head direction cell 
spiking structure: in full analogy with (\ref{sigmapath}), traversing a physical trajectory $s(t)$ induces a
sequence of active $\eta$-simplexes, or a head direction simplicial path
\begin{equation}
\tilde{\eta} =\{\eta_1,\eta_2,\ldots,\eta_n,\ldots\},
\label{etapath}
\tag{5$\eta$}
\end{equation}
in which different $\eta$-simplexes represent distinct directions, at all locations. As spiking information 
accumulates, the topological structure of $\mathcal{T}_{\eta}(t)$ converges to the structure of nerve complex
$\mathcal{N}_{\eta}$ induced by the head direction fields' cover of $S^1$---every $\eta$-simplex projects into 
its respective head direction field $\upsilon_{\eta}$ under the mapping (\ref{mapHD}). Simulations demonstrate
that in the environment shown on Fig.~\ref{fig:PFmap}A, a typical coactivity complex $\mathcal{T}_{\eta}(t)$ 
saturates in about $T_{\ast}^{\eta}\approx 2$ minutes, while the persistent homologies of $\mathcal{T}_{\eta}(t)$
filtered according to the times of $\eta$--simplexes' first appearances reveal the circular topology of the
space of directions in about $T_{\min}^{\eta}\approx 1.5$ minutes.

From the biological point of view however, these results do not provide an estimate for \textit{orientation 
learning time}: by itself, $T_{\min}^{\eta}$ may be viewed as the time required to learn head directions at a
particular location, in every environment, whereas learning to orient in $\mathcal{E}$ implies knowing directions
at every location and an ability to link orientations across locations. The latter is a much more extensive task,
which, as it will be argued below, requires additional specifications and interpretations.

The following discussion is dedicated to constructing phenomenological models of orientation learning using 
algebraic topology and synthetic geometry approaches. In Section~\ref{sec:topol}, we construct and test
a direct generalization of the topological model, similar to the one used in \cite{PLoS,Arai,Basso,Hoffman}
and demonstrate that it fails to produce biologically viable predictions for the learning period. 
In Section~\ref{sec:geom}, the topological approach is qualitatively generalized using an alternative scope
of ideas inspired by synthetic geometry. In Section~\ref{sec:learn}, it is demonstrated that the resulting 
framework allows incorporating additional neurophysiological mechanisms and acquiring the topological connectivity
of the environment in a biologically viable time, thus revealing a new level of organization of the cognitive map, 
as discussed in Section~\ref{sec:disc}.

\section{Topological model of orientation learning}
\label{sec:topol}

\textbf{Orientation coactivity complex}. The model requirements \textbf{R1} and \textbf{R2}, applied to both
hippocampal and head direction activity, imply that the animal's location and orientation are represented, at
any moment of time, by an active $\sigma$-simplex and an active $\eta$-simplex. Thus, the net pattern of activity
in the hippocampal and in the head direction networks defines a $(\sigma,\eta)$ pair---a single \textit{oriented},
or \textit{pose} simplex
\begin{equation}
\setcounter{equation}{6}
\zeta=[\sigma,\eta],
\nonumber
\end{equation}
(the latter term is borrowed from robotics \cite{Thrun,Heinze,Savelli}). Restricting a $\zeta$-simplex to its
maximal subsimplexes spanned, respectively, by the place- or the head direction cells defines the projections 
into its positional and directional components,
\begin{subequations}
	\label{proj}
	\begin{align}
	\pi_{\sigma}:\zeta\to \sigma,\label{projsigma}\tag{6$\sigma$}\\
	\pi_{\eta}:\zeta\to\eta,\label{projeta}\tag{6$\eta$}
	\end{align}
\end{subequations}
which permits terminology such as ``$\zeta$ is located at $\sigma$," ``$\zeta$ is directed toward $\eta$," ``a
location $\sigma$ is directed by $\eta$," ``$\eta$ is applied at $\sigma$," etc. Thus, one may refer to the
$\sigma$-simplexes as to \textit{locations} and to the $\eta$-simplexes as to \textit{directions}, implying,
depending on the context, either the items encoded in the cognitive map, or the $\sigma/\eta$-fields, or both.

As in the previously discussed cases, the collection of pose simplexes produced up to a moment $t$ forms an 
\textit{orientation coactivity complex} $\mathcal{T}_{\zeta}(t)$ that schematically represents the net pool of
conjunctive patterns generated by the place- and the head direction cells accumulated since the onset of the
navigation. In particular, the combinations of cells ignited along a physical path $s$ induces an 
\textit{oriented simplicial path}
\begin{equation}
\setcounter{equation}{7}
\tilde{\zeta}=\{\zeta_1,\zeta_2,\ldots,\zeta_n,\ldots\},
\label{zpath}
\end{equation} 
which runs through $\mathcal{T}_{\zeta}(t)$. The transitions from a given active pose simplex, $\zeta_i$,
to the next, $\zeta_{i+1}$, occur at discrete moments $t_1,t_2,\ldots,t_n,\ldots$, when either the $\sigma$- or
the $\eta$-component of $\zeta_i$ deactivates and the corresponding component of $\zeta_{i+1}$ ignites. Thus,
the simplicial paths $\tilde{\sigma}$ and $\tilde{\eta}$ can be produced from the oriented path (\ref{zpath})
using (\ref{proj}). In contrast with (\ref{sigmapath}) and (\ref{etapath}), the $\sigma$- and $\eta$-simplexes
in such paths are indexed uniformly, according to the indexes of (\ref{zpath}),
\begin{subequations}
	\label{pathz}
	\begin{align}
	\tilde{\sigma}=\{\sigma_1,\sigma_2,\ldots,\sigma_n\}, \label{psigma}\tag{8$\sigma$}\\
	\tilde{\eta}=\{\eta_1,\eta_2,\ldots,\eta_n\}.\label{peta}\tag{8$\eta$}
	\end{align}
\end{subequations}
The adjacent simplexes in either (\ref{psigma}) or in (\ref{peta}) (but not in both of them simultaneously)
may coincide, e.g., the location $\sigma_i$ may remain the same during several timesteps, while the $\eta$-activity
changes, or vice versa.

Since the rat can potentially run in any direction at any location (unless stopped by an obstacle), there are no 
\textit{a priori} restrictions on the order of the place cell and the head direction cells spiking activity. This 
observation is formalized by another model requirement: 

\vspace{2mm}
\textbf{R3. Independence}.
\textit{A given head direction cell assembly $\eta$ may become coactive with any place cell assembly $\sigma$ 
	and vice versa, with independent $\sigma$- and $\eta$-spiking parameters.}
\vspace{2mm}

In model's terms, this implies that the development of the coactivity complex $\mathcal{T}_{\eta}(t)$ and its
ultimate saturated structure $\mathcal{N}_{\eta}$ is the same at any location $\sigma$, and vice versa, the 
\textit{saturated} structure of $\mathcal{T}_{\sigma}(t)$ is independent from $\eta$-activity.

However, since the activities in the hippocampal and in the head direction cell networks represent complementary
aspects of the same movements, certain characteristics of the simplicial paths (\ref{pathz}) are coupled. 
Specifically, in light of \textbf{R1}--\textbf{R2}, a connected physical trajectory $s$ should induce a connected
$\sigma$-path in the place cell complex $\mathcal{T}_{\sigma}$, together with a connected $\eta$-path in the head
direction complex $\mathcal{T}_{\eta}$. Similarly, a \textit{looping} trajectory should induce periodic sequences
of simplexes,
\begin{eqnarray}
	\tilde{\sigma}_{o}=\{\sigma_1,\sigma_2,\ldots, \sigma_n, \sigma_1, \sigma_2,\ldots\},\nonumber\\
	\tilde{\eta}_{o}=\{\eta_1,\eta_2,\ldots,\eta_n, \eta_1, \eta_2, \ldots\}. \nonumber
\end{eqnarray}
In other words, making a loop in physical space $\mathcal{E}$ should induced $\sigma$- and $\eta$-loops. Thus, 
without referencing the physical trajectory, the model requires

\vspace{2mm}
\textbf{R4. Topological consistency}.
\textit{The simplicial paths (\ref{pathz}) should be connected and a simple periodic $\sigma$-path should 
	induce a simple periodic $\eta$-path and vice versa}.
\vspace{2mm}

\textbf{Orientation learning}. As discussed above, getting rid of the topological defects in 
$\mathcal{T}_{\sigma}(t)$ and in $\mathcal{T}_{\eta}(t)$ allows faithful topological classification of physical
routes in terms of the neuronal (co)activity. Thus, the ``topological maturation" of these complexes can be viewed as a
schematic representation of the learning process. The concept of oriented simplicial paths embedded into the 
orientation coactivity complex $\mathcal{T}_{\zeta}(t)$ allows a similar interpretation of the spatial orientation
learning---as acquiring an ability to distinguish between qualitatively disparate moving sequences. Indeed, knowing
how to orient in a given space, viewed as a cognitive ability to reach desired places from different directions
via a suitable selection of intermediate locations and turns, may be interpreted mathematically as an
ability to classify trajectories using topologically inequivalent classes of oriented simplicial paths (\ref{pathz}). 
From an algebraic-topological perspective, this may be possible after the orientation complex $\mathcal{T}_{\zeta}(t)$ 
acquires its correct topology.

\begin{figure}[h]
	\centering
	\includegraphics[scale=0.75]{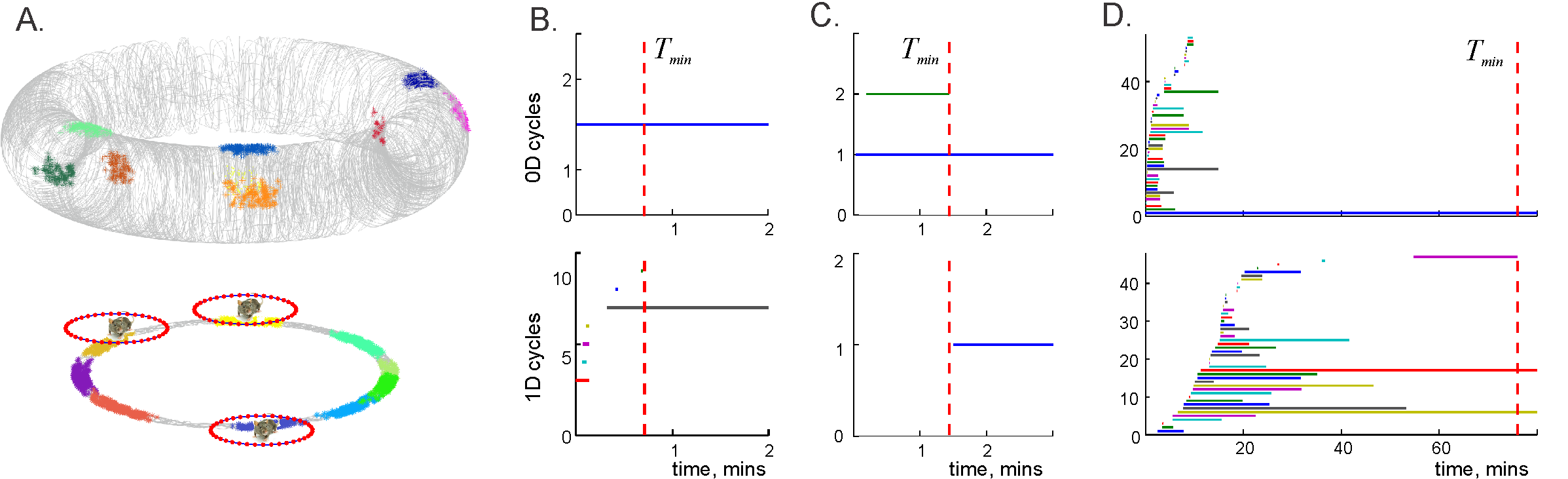}
	\caption{{\footnotesize\textbf{Topological leaning dynamics}. \textbf{A}. The orientation space for a rat
			moving on a circular runway (bottom) is a topological torus (top), $\mathcal{O}=T^2$. Clusters of 
			colored dots show examples of the simplex fields $\upsilon_{\zeta_{ij}}$ of the basic coactivity 
			combinations $\zeta_{ij}=[c_i,h_j]$. The timelines of the topological loops in the place cell complex 
			$\mathcal{T}_{\sigma}(t)$ (panel \textbf{B}, horizontal bars) and in the head direction cell complex 
			$\mathcal{T}_{\eta}(t)$	(panel \textbf{C}) disappear in under a minute; a stable loop in $0D$ and a 
			stable loop	in $1D$ in each case indicate that both the runway and the direction space are topological
			circles. 
			\textbf{D}.	The orientation coactivity complex $\mathcal{T}_{\zeta}(t)$ contains many topological
			defects that take over an hour to disappear: the spurious $0D$ loops contract in about $17$ minutes
			(top panel), and the spurious $1D$ loops persist for $~80$ minutes. For an ``open field" environment
			(Fig.~\ref{fig:PFmap}A) a similar topological learning process takes hours. These estimates exceed 
			the	experimental learning timescales, suggesting that the physiological orientation learning involves 
			additional mechanisms.
	}}
	\label{fig:Circ}
\end{figure}

To establish the latter, note that the complex $\mathcal{T}_{\zeta}(t)$ has the same nature as 
$\mathcal{T}_{\sigma}(t)$ and $\mathcal{T}_{\eta}(t)$---it is an emerging temporal representation of a nerve
complex, induced from a cover of a certain \textit{orientation space} $\mathcal{O}$ that combines $\mathcal{E}$
and $S^1$. Since the preferred angles of the head direction cells remain the same at all locations \cite{Wiener},
the space of directions $S^1$ represented by these cells does not ``twist'' as the rat moves across $\mathcal{E}$,
which implies that the  orientation space has a direct product structure $\mathcal{O}=\mathcal{E}\times S^1$ 
\cite{Hatcher}. One may thus combine (\ref{mapPC}) and (\ref{mapHD}) to construct a \textit{joint spatial mapping},
$f_{\zeta}=(f_{\sigma},f_{\eta})$, that associates instances of simultaneous activity of place- and head 
direction cell groups with domains in the orientation space, $$f_{\zeta}: \mathcal{T}_{\zeta}\to\mathcal{O}.$$
For example, if a given place cell $c$ maps into a field $\upsilon=f_{\sigma}(c)$, then the coactivity of a 
pair $\zeta=[c,h]$ (the smallest possible combined coactivity) can be mapped into $\mathcal{O}$ by shifting 
$\upsilon$ along the corresponding fiber $S^1$ according to the angle field of the head direction component 
of $\zeta$, $\varphi=f_{\eta}(h)$ (Fig.~\ref{fig:Circ}A). The resulting \textit{orientation fields}, 
$\upsilon_{\zeta_i}=f_{\zeta}(\zeta_i)$, form a cover of the orientation space, $$\mathcal{O}=\cup_i\upsilon_{\zeta_i},$$ 
whose nerve $\mathcal{N}_{\zeta}$ is reproduced by the temporal orientation complex $\mathcal{T}_{\zeta}(t)$. 

Just as the $\sigma$- and $\eta$-simplexes, each pose simplex $\zeta_k$ has a well-defined appearance time, 
$t_k$, due to which the orientation complex is \textit{time-filtered}, $\mathcal{T}_{\zeta}(t)\subseteq
\mathcal{T}_{\zeta}(t')$ for $t<t'$. 
Applying Persistent Homology techniques \cite{Zomorodian,Edelsbrunner}, one can compare topological shape
defined by the time-dependent Betti numbers of $\mathcal{T}_{\zeta}(t)$ with the shape of the orientation space
$\mathcal{O}=\mathcal{E}\times S^1$, and thus quantify the orientation learning process.

As an illustration of this approach, we simulated the rat's movements on a circular runway, for which the total
representing space for the combined place cell and head direction cell coactivity forms a $2D$ torus 
(Fig.~\ref{fig:Circ}A). To simplify modeling, we used movement direction as a proxy for the head direction, 
although physiologically these parameters not identical \cite{Cei,Raudies,Laurens,Shinder1,Shinder2}.
Computations show that the correct topological shapes of the place- and the head direction complexes emerge in
about $1$ minute (Fig.~\ref{fig:Circ}B,C), while the transient topological defects in $\mathcal{T}_{\zeta}(t)$
disappear in about $80$ minutes (Fig.~\ref{fig:Circ}D)---a surprisingly large value that exceeds behavioral 
outcomes by an order of magnitude \cite{Alvernhe,Piet}. Even assuming that biological learning may involve only
a partial reconstruction of the orientation space's topology, the persistence diagrams shown on Fig.~\ref{fig:Circ}D
indicate that $\mathcal{T}_{\zeta}(t)$ is not only riddled with holes for over $40$ minutes, but also that it
remains disconnected ($b_0(\mathcal{T}_{\zeta})>1$) for up until $17$ minutes, i.e., according to the model, 
the animal should not be able to acquire a connected map of a simple annulus after completing multiple lapses
across it.

\textbf{Biological implications} of mismatches between the experimental and the modeled estimates of learning
timescales are intriguing: since the model's quantifications are based on ``topological accounting" of place-
and head direction cell coactivities induced by the animal's movements, the root of the problem seems to lay
not in the mathematical side of the model, but in the biological assumptions that underlie the computations.
Specifically, the overly long learning periods suggest that building a spatial map from the movement-triggered 
neuronal coactivity alone does not take into account certain principal components of the learning mechanism. 
In other words, the fact that the animal seems to produce correct representations of the environment much faster
than it would be possible from the influx of navigation-triggered data, implies that the brain can bypass the
necessity to discover every bit of information empirically, i.e., that building a cognitive map may be 
accelerated by ``generating information from within,'' via autonomous network dynamics.

Physiologically, this conclusion is not surprising: the phenomena associated with spatial information processing
through endogenous hippocampal activity are well known. Many experiments have demonstrated that the animal can 
\textit{replay} place cells during the quiescent states \cite{Wu,Karlsson1,Olafsdottir} or sleep \cite{Ji,Louie},
in the order in which they have fired during preceding active exploration of the environment, or \textit{preplay}
place cells in sequences that represent future trajectories \cite{Pfeiffer1,Johnson,Dragoi}. 
These phenomena are commonly viewed as manifestations of the animal's ``mental explorations'' of its cognitive
map, which help acquiring, sustaining and retrieving memories \cite{Hopfield,Zeithamova}.

However, one would expect a different functional impact of replays and preplays on spatial learning. Since 
replays represent past experiences, they cannot accelerate acquisition of new spatial information---in the 
model's terms, reactivation of simplexes that are already included into the coactivity complex cannot alter
its shape (in absence of synaptic and structural instabilities \cite{Roux,Ven,PLoZ,Replays}). In contrast,
preplaying place cell combinations that have not yet been triggered by previous physical moves may speed up
the learning process. Yet, from the modeling perspective, it is \textit{a priori} unclear which specific
trajectories may be preplayed by the brain, or, in computational terms, which specific simplicial trajectories
should be ``injected" into the simulated cognitive map $\mathcal{T}_{\zeta}(t)$ to simulate the preplays that
may accelerate learning. Experiments suggest that ``natural" connections between locations are the straight
runs \cite{Pfeiffer1,Valerio,McNPth}; however, implementing such runs would require a certain ``geometrization"
of the topological model. In the following, we propose a geometric implementation of preplays that help to
expedite learning process and open new perspectives modeling spatial representations.

\section{Geometric model of orientation learning}
\label{sec:geom}

\textbf{The motivation} for an alternative approach comes from the observation that combining inputs from the
place and the head direction cells offers a possibility of establishing different arrangements of the locations 
in the hippocampal map. Indeed, common interpretations of the head direction cells' functions suggest that the
rat's movements guided by a fixed head direction activity trace approximately straight paths, whereas shifts in
$\eta$-activity indicate curved segments of the trajectory, turns, etc. \cite{ChenMcN,Wiener,Savelli}. 
This implies that the brain may use the head direction cells' outputs to represent shapes of the paths encoded
by the place cells, to align their segments, identify collinearities, their incidences, parallelness, etc.

To address these structures and their properties we will use the following definitions:
\vspace{2 mm}

\textbf{D1}.\label{def:ald}
Simplexes $\sigma_1$ and $\sigma_2$ are \textit{aligned in $\eta$-direction}, if they may ignite during an
uninterrupted activity of a fixed $\eta$-simplex. In formal notations, $(\sigma_1,\sigma_2)\lhd\eta$.
 
\textbf{D2}.\label{def:adj} 
Two $\eta$-aligned simplexes are $\eta$-\textit{adjacent}, $\{\sigma_1\multimapdot\sigma_2|\eta\}$, if the
ignition of $\sigma_2$ follows immediately the ignition of $\sigma_1$, with no other cell groups igniting in-between.

\textbf{D3}.\label{def:alm} 
An ordered sequence of $\sigma$-simplexes forms an \textit{$\eta$-oriented alignment} if each pair of consecutive
simplexes, $(\sigma_i,\sigma_{i+1})$ in (\ref{psigma}), is $\eta$-adjacent, i.e., if the oriented path,
$$\ell=\{[\sigma_1,\eta],[\sigma_2,\eta],\ldots,[\sigma_n,\eta]\},$$ never changes direction. The notation
\begin{equation}
\ell=\{\sigma_1,\sigma_2,\ldots,\sigma_n|\eta\}\equiv\{\bar{\sigma}|\eta\}
\label{align}
\end{equation} 
highlights the set of \textit{collinear} locations $\bar{\sigma}=(\sigma_1,\sigma_2,\ldots,\sigma_n)$
and the $\eta$-simplex that orients it. The bar in $\bar{\sigma}$ is used to distinguish an alignment from a
generic simplicial path $\tilde{\sigma}$.

\textbf{D4}.\label{def:Ex}
An alignment $\ell_1$ \textit{augments} an alignment $\ell_2$, if both $\ell_1$ and $\ell_2$ can be guided 
by an uninterrupted $\eta$-activity 
($\bar{\sigma}_1\Join\bar{\sigma}_2\Leftrightarrow(\bar{\sigma}_1\cup\bar{\sigma}_2)\lhd\eta$). Conversely, a
proper subset $\bar{\sigma}'$ of an aligned set $\bar{\sigma}$ forms its proper \textit{subalignment}
($\bar{\sigma}'\subset\bar{\sigma}\Leftrightarrow\{\bar{\sigma}'|\eta\}\ltimes\{\bar{\sigma}|\eta\}$).

\textbf{D5}.\label{def:over} 
Two alignments $\ell_1$ and $\ell_2$ \textit{overlap}, if they share a location $\sigma$ ($\ell_1\cap\ell_2=
\sigma\Leftrightarrow\sigma\in\bar{\sigma}_1\cap\bar{\sigma}_2$).

\textbf{D6}.\label{def:out} 
A location $\sigma'$ lays \textit{outside} of an $\eta$-alignment $\ell_{\eta}$, if it aligns with any $\sigma$
from $\ell_{\eta}$ along a direction different from $\eta$
($\sigma'\notin\ell_{\eta}\Leftrightarrow\exists\sigma\in\ell_{\eta},\eta'\neq\eta:(\sigma,\sigma')\lhd\eta'$).

\textbf{D7}.\label{def:Prla}
Two alignments are \textit{parallel}, if they are directed by the same or opposite $\eta$-activity, without
augmenting each other, i.e., if one $\pm\eta$-alignment contains a location outside of the other one, ($\ell_1
\parallel\ell_2\Leftrightarrow \eta_1 =\pm\eta_2$, and $\exists\sigma:\sigma\in\bar{\sigma}_1\, ,\sigma\notin
\bar{\sigma}_2$, where $f_{\eta}(-\eta)\approxeq\pi+f_{\eta}(\eta)$).

\textbf{D8}.\label{def:yaw}
A \textit{yaw} is an oriented path in which a sequence of $\eta$-simplexes ignites at a fixed location $\sigma$,
$$\vary=\{[\sigma,\eta_1],[\sigma,\eta_2],\ldots,[\sigma,\eta_n]\}.$$ Thus, yaws may be viewed as structural 
opposites of the alignments, which is emphasized by the notation
\begin{equation}
\vary=\{\eta_1,\eta_2,\ldots,\eta_n|\sigma\}\equiv\{\hat{\eta}|\sigma\}
\nonumber
\end{equation}
that highlights the range of $\eta$-simplexes, $\hat{\eta}$, ignited at the \textit{axis of the yaw}, $\sigma$. 

\textbf{D9}.\label{def:turn} 
A \textit{clockwise turn} is an oriented path $\tilde{\zeta}_{+}=\{[\sigma_1,\eta_1],[\sigma_2,\eta_2],\ldots,
[\sigma_n,\eta_n]\}$ with a growing angular sequence, i.e., the angle $\varphi_i$ that represents the element
$\eta_i$ is not greater than the next one, $\varphi_{i+1}\geq\varphi_i$. A \textit{counterclockwise turn} 
$\tilde{\zeta}_{-}$ is an oriented path with a decreasing angular sequence, $\varphi_{i+1}\leq\varphi_i$.

The latter definition is due to the observation that $\eta$-simplexes can be ordered according to the angles
they represent, i.e., $\eta_i<\eta_{i+1}$ iff $\varphi_i<\varphi_{i+1}$, which also allows defining the 
angle between alignments, 
\begin{equation}
\measuredangle(\ell_{\eta_i},\ell_{\eta_j})\equiv\measuredangle(\eta_i,\eta_j)\equiv|\varphi_i-\varphi_{j}|=
\Delta\varphi_{ij}.
\nonumber
\end{equation}

\vspace{2 mm}

In light of the definitions \textbf{D1}--\textbf{D9}, the model requirements \textbf{R1}--\textbf{R4} imply that
the entire ensembles of the active place and head direction cells are involved into geometric arrangements, e.g., 
every location $\sigma$ belongs to an alignment directed by a $\eta$-simplex, and conversely, every $\eta$-simplex  
directs a nonempty alignment through the $\sigma$-map. The question is, whether this collection of alignments is
sufficiently complete to allow self-contained geometric reasoning in terms of collinearities, incidences, parallelisms, 
etc., i.e., does it form a self-contained geometry? 

The standard approach to answering this question is based on verifying a set of axioms, in this case---the
axioms of affine geometry, applied to the elements of a suitable set $A=\{x,y,z,\ldots\}$ and its select subsets 
$\ell_1,\ell_2,\ldots,\in A$:

\vspace{2 mm}
\textbf{A1}. \textit{Any pair of distinct elements of $x\neq y$ is included into a unique subset $\ell$}.

\textbf{A2}. \textit{There exists an element $x$ outside of any given subset $\ell$, $x\notin \ell$}.

\textbf{A3}. \textit{For any subset $\ell$ and an element $x\notin \ell$, there exists a unique subset $\ell_x$
	that includes $x$, but does not overlap with $\ell$}.
\vspace{2 mm}

If these axioms (referred to as \textit{\textbf{A}-axioms} below) are satisfied, then the subsets $\ell_1,\ell_2,
\ldots$, can be viewed as lines because interrelationships among them and with other elements of $A$ reproduce
the familiar geometric incidences between lines and points in the Euclidean plane \cite{Hilbert,Vossen,Batten,Karteszi}. 
However, the set of geometries established via the \textbf{A}-axioms is much broader than its main ``motivating 
example:" the standard planar affine geometry $\mathscr{A}_E$ is but a specific model implementing the 
\textbf{A}-axioms using the infinite set of infinitesimal points and infinite lines \cite{Hilbert,Vossen,Batten,Karteszi}. 
In fact, it is also possible to use the \textbf{A}-axioms to establish geometry on finite sets, thus producing
finite affine planes. This is important for modeling cognitive maps encoded by the physiological networks that 
contain finite numbers of neurons and thus may represent finite sets of locations and alignments. Specifically,
a possible adaptation of the \textbf{A}-axioms using spiking semantics, is the following: 

\vspace{2 mm}

\textbf{A1}$_n$. \textit{Any pair of distinct locations $\sigma_1\neq \sigma_2$ belongs to a unique alignment, 
i.e., $\sigma_1$ and $\sigma_2$ may ignite in sequence during the activity of a single $\eta$-simplex}
($\forall\sigma_1,\sigma_2,\exists!\eta, \bar{\sigma}:(\sigma_1,\sigma_2)\ltimes\bar{\sigma}\lhd\eta$).

\textbf{A2}$_n$. \textit{The location-encoding network can activate a group of cells $\sigma$ to represent a
	location outside of any given alignment} ($\forall\ell,\,\exists\sigma\notin\ell$).

\textbf{A3}$_n$. \textit{For any alignment $\ell$ and a location $\sigma\notin\ell$, there exists a unique 
	alignment $\ell_{\sigma}$ parallel to $\ell$ that passes through $\sigma$} 
($\forall\ell,\,\sigma\notin\ell,\,\exists!\ell_{\sigma}: \sigma\in\ell_{\sigma}, \ell\parallel\ell_{\sigma}$).
\vspace{2 mm}

\begin{wrapfigure}{c}{5 cm}
	\includegraphics[scale=1.16]{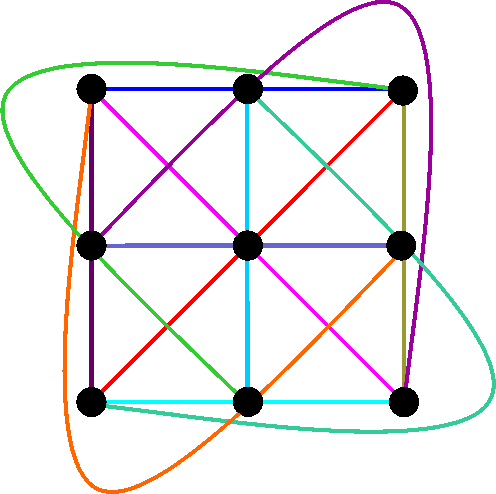}
	\caption{
		{\footnotesize \textbf{Finite affine plane} of the third order, $\mathscr{A}_3$, with $9$ points 
			(black dots) aligned according to the \textbf{A}-axioms in $12$ lines (colors) forms a non-planar
			configuration.}
	}\label{fig:A3} 
\end{wrapfigure}

Validating these axioms over the net pool of spiking activities produced by the hippocampal and head direction
cells would establish a discrete-geometric structure encoded by $(\sigma,\eta)$-neuronal activity. However,
the requirements imposed by the \textbf{A}$_n$-axioms may not be compatible with physiological mechanisms that 
operate the corresponding networks, as well as with these networks' functions. Indeed, the configurations formed
by the connections in finite planes are typically non-planar (Fig.~\ref{fig:A3}), whereas physiological 
computations combining place and head direction cells' activities appear to enable geometric planning in planar
environments \cite{Valerio,McNPth}. Second, the combinations of locations that form ``relational lines" according
to the \textbf{A}$_n$-axioms may not have the standard properties of their Euclidean counterparts, e.g., they may
include sequences of locations that cannot be consistently mapped into straight Euclidean paths (Fig.~\ref{fig:A3}).
In contrast, experiments show that neuronal activity during animals' movements along straight arrangements of
$\sigma$-fields, as well as their offline preplays/replays \cite{Mattar,Byrne}, dovetail with the definitions
\textbf{D1}-\textbf{D7}. Third, given a large number of the encoded locations (in rats, about $3\times 10^4$ of
active place cells in small environments \cite{Ziv}) and a very large set of possible co-active cell combinations
\cite{Syntax,CAs}, a finite set of $\eta$-simplexes may not suffice to align all pairs of locations in the sense
of the definition \textbf{D7}.
 
On the other hand, certain key features of finite affine planes, e.g., the necessity of having $k$ parallel lines
in every direction and a fixed number, $k+1$, of lines passing through each location \cite{Hilbert,Vossen,Batten,Karteszi}
are reflected in the network. Indeed, the existence of a fixed population of head direction assemblies results
in a fixed number $k\approxeq N_{\eta}/2$ of distinct alignments passing through any location and the same number
of directions running across the cognitive map.

Together, these observations suggest that the brain may combine certain aspects of finite geometries and 
Euclidean plane discretizations. Rather than trying to recognize the net geometry of the resulting 
representations from the onset, one may adopt a constructive ``bottom up" approach: it may be possible to
interpret certain \textit{local} properties of spiking activity as basic geometric relations and then follow
how such relations accumulate at larger scales, yielding \textit{global} geometric frameworks.
For example, it can be argued that place cell activities can be aligned locally, i.e., that ignitions of a
specific head direction assembly can accompany transitions of activity from one place cell assembly to an 
adjoining one. It is also plausible that, in stable network configurations, the selection of cell groups
involved in such transitions is limited or even unambiguous. Also, given the number of place cells
($N_c\approx 3\times 10^5$) and typical assembly sizes ($60-300$ cells) \cite{Syntax}, there should be
enough place cell combinations to represent a sufficient set of alignments, overlaps between them, etc. 
\cite{CAs,Perin,Reimann}.

Thus, in addition to (mostly topological) requirements \textbf{R1}--\textbf{R4},
one can consider the following neuro-geometric rules:

\vspace{2pt}

\textbf{G1}. \textit{Any two adjacent locations align in a unique direction} $(\forall\sigma_1\multimapdot\sigma_2,
\,\exists!\eta: (\sigma_1,\sigma_2)\lhd\eta)$.

\textbf{G2}. \textit{A location adjacent to a given one may be recruited in any direction}
			 $(\forall\eta,\sigma_1,\tilde{\exists}!\sigma_2:(\sigma_1\multimapdot\sigma_2)\lhd\eta$).
			 
\textbf{G3}. \textit{The location-encoding network can explicitly represent the overlap between any two non-parallel
	alignments}
($\forall\ell_1,\ell_2,\eta_1\neq\eta_2,\,\tilde{\exists}!\sigma:\ell_1\cap\ell_2=\sigma\in\ell_1,\ell_2$).

\vspace{2pt}

In contrast with the \textbf{A}$_n$-axioms that aim to establish large-scale properties of a ``cognitive"
affine plane as a whole, the \textbf{G}-rules define local geometric relationships induced by local mechanisms
controlling neuronal activity. In particular, \textbf{G1} ascertains a possibility of aligning any two adjacent
locations, rather than any two locations as required by the \textbf{A1}$_n$. The rule \textbf{G2} is complementary: it
posits that if an active $\eta$-combination is selected, then the activity can propagate from a given $\sigma_1$
to a specific adjacent $\sigma_2$. Lastly, the rule \textbf{G3} allows reasoning about the locations, alignments,
incidences, etc., assuming that all these elements can be physiologically actualized.

\begin{figure}[h!]
	\centering
	\includegraphics[scale=0.84]{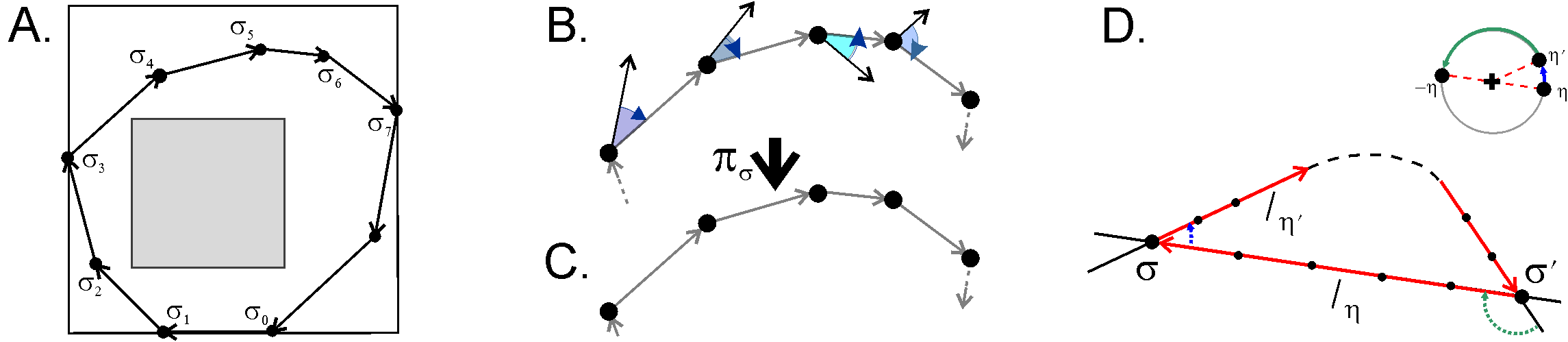}
	\caption{
		{\footnotesize \textbf{G-rules based geometric constructions}.
			\textbf{A}. A directed polygonal chain connecting pairs of adjacent locations along the navigated
			trajectory in the environment shown on Fig.~\ref{fig:PFmap}A. 
			\textbf{B}. A schematic representation of a discrete homotopy from a generic $\zeta$-path to a 
			polygonal chain: the $\eta$-components shift towards the unique directions that align the adjacent 
			locations (gray arrows).
			\textbf{C}. The resulting polygonal chain (a combination of yaws and straight runs) projects by
			$\pi_{\sigma}$ into a undirected chain connecting the adjacent locations---a fragment of the chain
			shown on the panel \textbf{A}.
			\textbf{D}. Lemma 2: If two nonparallel alignments, $\ell_{\eta}$ and $\ell_{\eta'}$, produce two
			intersections $\sigma$ and $\sigma'$, then there exists a $\tilde{\sigma}$-trajectory that forms a 
			noncontractible simple loop, while the corresponding $\tilde{\eta}$-trajectory forms a contractible
			segment (top left corner), in contradiction with \textbf{R4}.}
	}\label{fig:polig} 
\end{figure}

As a first application of the \textbf{G}-rules, notice that a $\sigma$-path, viewed as a sequence of adjacent 
$\sigma$s, induces a unique ordered $\eta$-sequence, i.e., a $\tilde{\eta}$-path in $\mathcal{T}_{\eta}(t)$.
Together, these paths define an oriented trajectory $\tilde{\zeta}'$, formed by uniquely directed straight
links between adjacent locations, which can be graphically represented by a \textit{directed polygonal chain}
of $\sigma$-locations (Fig.~\ref{fig:polig}A). Conversely, the fact that a generic $\tilde{\zeta}$ projects 
into a ordered sequence of adjacent $\sigma$-simplexes, $\tilde{\sigma}=\pi_{\sigma}(\tilde{\zeta})$, implies
that oriented paths can be aligned into the polygonal chains, $\tilde{\zeta}\to\tilde{\zeta}'$ 
(Fig.~\ref{fig:polig}B). From the perspective of the topological model discussed in Section~\ref{sec:topol}, 
this means that each path $\tilde{\sigma}$ can be ``lifted" from $\mathcal{T}_{\sigma}(t)$ into a unique oriented
path $\tilde{\zeta}'\in\mathcal{T}_{\zeta}(t)$ by a back projection, $\tilde{\zeta}'=\pi_{\sigma}^{-1}(\tilde{\sigma})$.
In the following, the term ``simplicial path" will refer to the polygonal chains only, unless explicitly stated 
otherwise, and the ``prime" notation will be suppressed.

Reversing the order of simplexes in a chain and inverting the corresponding $\eta$-sequence,
\begin{subequations}
	\label{rever}
	\begin{align}
	\tilde{\sigma}_{+}=\{\sigma_{i_1},\sigma_{i_2},\ldots,\sigma_{i_n}\}\to\tilde{\sigma}_{-}=
	\{\sigma_{i_n},\sigma_{i_{n-1}},\ldots,\sigma_{i_1}\}, \label{rsigma}\tag{10$\sigma$}\\
	\tilde{\eta}_{+}=\{\eta_{i_1},\eta_{i_2},\ldots,\eta_{i_n}\}\to \tilde{\eta}_{-}=
	\{\pm\eta_{i_n},\pm\eta_{i_{n-1}},\ldots,\pm\eta_{i_1}\}.\label{reta}\tag{10$\eta$}
	\end{align}
\end{subequations}
where the angle $-\eta$ is diametrically opposite to $\eta$, $f_{\eta}(-\eta_i)\approxeq\pi+f_{\eta}(\eta_i)$.
The ``$+$" sign in (\ref{reta}) corresponds to ``backing up" along the path $\tilde{\sigma}_{+}$ and the ``$-$"
sign to reversing the moving direction, either by implementing the required physical steps or by flipping the
order of the replayed or preplayed sequences \cite{Foster,Ambrose} (in open fields, place cell spiking is 
omnidirectional \cite{ChenG1}). The Since move reversal does not affect the $\sigma$-paths' geometries, the
transformations (\ref{rever}) can be regarded as equivalence relationships, which do not reference physical
trajectory:

\vspace{1 mm}
\textbf{R5. Reversibility}.
\textit{Simplicial $\sigma$-paths related via (\ref{rever}) are geometrically identical.}
\vspace{1 mm}

In accordance with \textbf{R5}, a given $\eta$-oriented alignment, $\ell_{+}=\{\bar{\sigma}|\eta\}$, and its
inverse, $\ell_{-}=\{\bar{\sigma}|-\eta\}$, define the same collinear sequence, i.e., $\pi_{\sigma}(\ell_{+})=
\pi_{\sigma}(\ell_{-})=\bar{\sigma}$---a natural observation that motivates the definition \textbf{D7}. 
In particular, a pair of $\pm\eta$ adjacent locations is also geometrically adjacent ($\sigma_i\multimapdotboth
\sigma_j\Leftrightarrow\{\sigma_i\multimapdot\sigma_j|\eta\}\,\textrm{ or }\,\{\sigma_i\multimapdotinv\sigma_j|
-\eta\}$), which allows representing trajectories by \textit{undirected polygonal chains} connecting adjacent
$\sigma$-fields (Fig.~\ref{fig:polig}C). For example, a bending chain corresponds to a clockwise turn 
$\tilde{\zeta}_{+}$ as well as to its counterclockwise counterpart $\tilde{\zeta}_{-}$ (both project to the 
same $\sigma$-path, $\pi_{\sigma}(\tilde{\zeta}_{+})=\pi_{\sigma}(\tilde{\zeta}_{-})=\tilde{\sigma}$); a 
closed chain---to a loop that can be traversed in clockwise or in counterclockwise direction 
($\tilde{\sigma}_{o}=\pi_{\sigma}(\tilde{\zeta}_{o_{-}})=\pi_{\sigma}(\tilde{\zeta}_{o_{+}})$), etc. 

Returning to the link between the \textbf{G}-rules and the remaining two \textbf{A}- or \textbf{A}$_n$-axioms,
it can be observed that the \textbf{A2}$_n$-axiom is an immediate consequence of \textbf{G2}: if the network 
is capable of actualizing up to $N_{\eta}$ simplexes adjacent to a given one along $N_{\eta}$ available directions, 
then $N_{\eta}-2$ of them will necessarily lay outside of a given alignment. The argument for the existence and
uniqueness of parallel lines (axiom \textbf{A3}$_n$) can be organized into the following two lemmas:

\vspace{1 mm}
\textbf{Lemma 1}. \textit{If $\sigma$ is a location outside of an alignment $\ell_{\eta}$, $\sigma\notin\ell_{\eta}$, 
and $\ell_{\eta}'$ is an alignment directed by $\eta$ at $\sigma$, $\ell_{\eta}\neq\ell'_{\eta}$, then $\ell_{\eta}$
and $\ell_{\eta}'$ do not overlap.}
\vspace{1 mm}

\textbf{Proof}. 
Assume that the overlap exists, $\ell_{\eta}\cap\ell_{\eta}'=\sigma'\neq\varnothing$. Since $\eta$ is a unique
index of directions, the location $\sigma'$ is $\eta$-aligned with its adjacent locations both in $\ell_{\eta}$
and in $\ell_{\eta'}$. Thus, $\ell_{\eta}$ and $\ell_{\eta}'$ augment each other ($\bar{\sigma}_1\Join 
\bar{\sigma}_2$), forming a single joint $\eta$-alignment that passes through $\sigma$, in contradiction with
the original assumption $\sigma\notin\ell_{\eta}$. $\blacksquare$ 

\vspace{2 mm}
\textbf{Lemma 2}. \textit{Two non-parallel lines cannot intersect more than once.}

\textbf{Proof}. 
Consider two alignments $\ell_{\eta}$ and $\ell_{\eta'}$, $\eta\neq\eta'$, with $\ell_{\eta}\cap\ell_{\eta'} 
= \sigma$. Without loss of generality (change $\ell_{\eta}\to\ell_{-\eta}$ if necessary), we may assume that 
the angle between them is sharp, $\measuredangle(\ell_{\eta},\ell_{\eta'})<\pi/2$ (Fig.~\ref{fig:polig}C).
Consider an oriented path $\zeta$ that starts at $\sigma$ in $\eta'$-direction, i.e., along $\ell_{\eta'}$. 
If $\ell_{\eta'}$ crosses $\ell_{\eta}$ again at a location $\sigma'$, then the path $\zeta$ may turn back at
$\sigma'$ and continue along $\ell_{-\eta}$ towards $\sigma$, then continue along $\ell_{\eta'}$ again, etc.,
yielding a single closed $\sigma$-path. On the other hand, the corresponding $\eta$-path links $\eta$ and 
$\eta'$ at the first turn and then goes back from $\eta'$ to $-\eta$ at the second turn, forming a contractible
segment, in contradiction with \textbf{R4}. $\blacksquare$

\vspace{2 mm}
In effect, these two lemmas validate the constructive definition of parallelness \textbf{D7} and point at an 
alternative form of the axiom \textbf{A3}$_n$: \textit{If two locations $\sigma_1$ and $\sigma_2$ are 
	aligned along $\eta$, then any two alignments $\ell_1$	and $\ell_2$ directed through $\sigma_1$ and 
	$\sigma_2$ by a $\eta'\neq\eta$ are parallel}.

The foregoing discussion suggests that the spatial framework represented by the place cells and the head
direction cells forms neither a na\"ive discretization of the Euclidean plane nor a conventional finite
geometry, as defined by the standard \textbf{A}-axioms. Rather, combining the location and the direction 
information can be used to capture a certain sub-collection of geometric arrangements, e.g., a particular 
set of alignments, which may then be used for navigation and geometric planning \cite{Valerio,McNPth}. 
Correspondingly, orientation learning can be interpreted as a process of establishing and expanding such
arrangements (e.g., prolonging shorter alignments, completing partial ones, etc.) and accumulating them in
the cognitive map.

\section{Synthesizing cognitive geometry}
\label{sec:learn}

\begin{figure}[h!]
	\centering
	\includegraphics[scale=0.86]{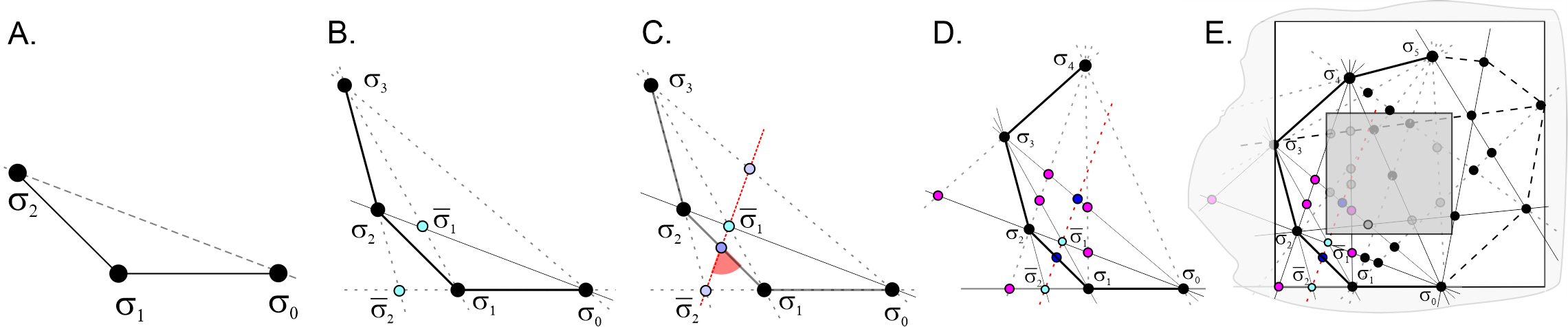}
	\caption{{\footnotesize\textbf{Aligning the cognitive map}.
			\textbf{A}. The endpoints of the initial two segments of the trajectory are connected by a home 
			run preplay $\ell_{20}$ (dashed line). 
			\textbf{B}. Reaching the next location, $\sigma_3$, allows preplaying home runs to $\sigma_0$
			and $\sigma_1$ and introducing the intersections $\bar{\sigma}_1$ and $\bar{\sigma}_2$ into the map.
			\textbf{C}. The direction of the alignment connecting the new points $\bar{\sigma}_1$ and $\bar{\sigma}_2$
			remain undefined, so $\ell_x=\{\bar{\sigma}_1,\bar{\sigma}_2|\eta_x\}$ and may thus be viewed as 
			provisional (topological) alignment in $\mathscr{A}_{\mathcal{C}}(t_3)$.
			\textbf{D}. The location $\sigma_4$ induces at several additional preplays to previously visited 
			locations. 
			\textbf{E}. At each step $t_n$, the adjacent segments of the trajectory, $\sigma_i\multimapdotboth\sigma_j$ 
			induce a \textit{connectivity graph} $\mathcal{G}_{\sigma}(t_n)$, and the corresponding clique complex 
			$\mathcal{T}_{\sigma}^{\ell}(\mathcal{G}_{\sigma})$ schematically represents the topological structure 
			of the aligned map $\mathscr{A}_{\mathcal{C}}(t_n)$. 
			\textbf{F}. The decays of the cell assemblies representing the unvisited locations (shaded areas) 
			induces the required topological dynamics.
	}}
   \label{fig:ind}
\end{figure}

As a basic example of a geometric map learning, consider an oriented trajectory $\zeta(t)$ that starts with an
alignment $\ell_1=\{\sigma_0,\sigma_1|\eta_1\}$, followed by a yaw at $\sigma_1$, and continues along $\ell_2=
\{\sigma_1,\sigma_2|\eta_2\}$, reaching $\sigma_2$ at the moment $t_2$ (Fig.~\ref{fig:ind}A). As in Sec.~\ref{sec:topol}, the
head and the motion directions are identified for simplicity. If $\sigma_0$, $\sigma_1$
and $\sigma_2$ are the only locations in the emerging affine map $\mathscr{A}(t_2)$, then $\sigma_2$ is adjacent to 
$\sigma_0$ and hence it must align with $\sigma_0$ along a certain $\eta$-direction $\eta_{20}$ (assuming a 
generic case, in which $\ell_1$ and $\ell_2$ are nonparallel, $\eta_1\neq\pm\eta_2$). Representing this alignment
in the parahippocampal network, i.e., producing the corresponding imprints in the synaptic architecture via 
plasticity mechanisms \cite{Leuner,Caroni,HebbRev}, requires actualization by igniting $\sigma_2$ and $\sigma_0$
consecutively during the activity of a particular $\eta_{20}$.
This can be achieved either by navigating between the corresponding $\sigma$-fields or off-line, via autonomous
network activity. In the former case, the connection $\ell_{20}=\{\sigma_2,\sigma_0|\eta_{20}\}$ is incorporated
into the map after the animal arrives to $\sigma_0$ from $\sigma_2$, i.e., at the ``empirical learning" timescale
discussed in Section~\ref{sec:topol}. In the latter case, $\ell_{20}$ may form at the spontaneous spiking
activity timescale (milliseconds \cite{Wu,Karlsson1,Olafsdottir,Ji,Louie,Pfeiffer1,Johnson,Dragoi}), as soon as
the animal reaches $\sigma_2$, which clearly accelerates the formation of the cognitive map.

This illustrates the model's general approach: although the geometric constructions were discussed in 
Section~\ref{sec:geom} in reference to spiking produced during the animal's movements, they also apply to endogenous
spiking activity. In other words, the \textbf{G}-rules can be used ``imperatively," for producing geometric 
structures in the cognitive map autonomously, based on available information rather than physical navigation.
In particular, preplays can be used for aligning locations with 
specific $\eta$-assemblies by preplaying straight ``home runs,'' as soon as the physical trajectory assumes a 
suitable configuration.

The physiological processes that enforce transitions of activity between cell assemblies are currently studied
both experimentally and theoretically \cite{Harris,Syntax,Laurens}; in case of the head direction and place cells,
the corresponding network computations may be guided by sensory (e.g., visual) and idiothetic (proprioceptive, 
vestibular and motor) inputs and involve a variety neurophysiological mechanisms \cite{Haggerty,KnierimId,ChenG2,ZhangS,Laurens}.
However, the principles of utilizing such mechanisms for acquiring a map of orientations can be illustrated using 
basic, self-contained algorithms that rely on the information provided by the hippocampal and head direction 
spiking. Specifically, the history of the $\eta$-assemblies' ignitions allows estimating the direction between
the loci of a polygonal chain according to
\begin{equation}
\Delta\varphi_{ij}=\frac{1}{N}\sum_{k=i}^j \varphi_k n_k,
\nonumber
\end{equation}
where $n_k$ is the number of spikes produced by the $k^{\textrm{th}}$ head direction assembly $\eta_k$, $\varphi_k$
is the corresponding angle (i.e., $\eta_k=\pi_{\eta}(\zeta_k)$ and $\varphi_k=f_{\eta}(\eta_k)$), $N$ is the total 
number of spikes. If $\eta_k$ is characterized by a Poisson firing rate $\mu_k$, then the number of spikes that it
produces over an ignition period $\Delta t_k$ can be estimated as $n_k=\mu_k \Delta t_k$. Assuming for simplicity
that all rates are the same $\mu_k=\mu$, the angular shifts can be estimated from the individual ignitions' duration
and the total navigation time $T$,
\begin{equation}
\Delta\varphi_{ij}=\frac{1}{T}\sum_{k=i}^j \varphi_k \Delta t_k.
\label{fitime}
\end{equation}
The $\eta$-simplex required to perform a home run from $\sigma_j$ to $\sigma_i$ can then be selected as the one
whose discrete angle is closest to $\varphi_j=\varphi_i+\Delta\varphi_{ij}$, i.e.,
\begin{equation}
\eta_j=\min_{\eta}\left(\varphi_j-f_{\eta}(\eta)\right).
\label{etasimplex}
\end{equation}
In particular, (\ref{fitime}) and (\ref{etasimplex}) allow estimating the required direction from $\sigma_2$
to $\sigma_0$ and thus identifying the simplex $\eta_{20}$ that needs to direct the corresponding home run
preplay $\ell_{20}$. Other models can be built by modifying or altering these rules.

The next move continues along $\ell_3=\{\sigma_2,\sigma_3|\eta_3\}$, arriving to $\sigma_3$ at the moment
$t_3$, which allows preplaying connections to previously visited locations along $\ell_{31}$ and $\ell_{30}$
in the map $\mathscr{A}_{\mathcal{C}}(t_3)$ (Fig.~\ref{fig:ind}B). If $\zeta(t)$ is a right turn ($\eta_1<\eta_2<\eta_3$),
then the line $\ell_{31}$ lays \textit{between} $\ell_{30}$ and $\ell_{3}$, and, according to \textbf{G3}, 
overlaps with $\ell_{20}$ at $\bar{\sigma}_1$, which will thus lay between $\sigma_0$ and $\sigma_2$, $\ell_{31}
=\{\sigma_3, \bar{\sigma}_1,\sigma_1|\eta_{31}\}$. Also, since $\ell_3$ and $\ell_1$ are non-parallel, they 
produce an overlap at $\bar{\sigma}_2$ that extends the ``seed alignments" $\ell_1(t_2)$ and $\ell_3(t_2)$ to
$\ell_1(t_3) = \{\sigma_0,\sigma_1,\bar{\sigma}_2|\eta_1\}$ and $\ell_3(t_3)=\{\bar{\sigma}_2,\sigma_2,\sigma_3|\eta_3\}$. 
The locations within the alignments $\ell_1$ and $\ell_3$ are ordered correspondingly, e.g., $\sigma_1$ falls
between $\sigma_0$ and $\bar{\sigma}_2$, and $\sigma_2$ falls between $\bar{\sigma}_2$ and $\sigma_3$. 

Note that since $\bar{\sigma}_1$ and $\bar{\sigma}_2$ can be viewed as adjacent, it is also possible to form 
an additional alignment $\ell_x=\{\bar{\sigma}_1,\bar{\sigma}_2|\eta_x\}$, which induces two additional 
locations by intersecting $\ell_{30}$ and $\ell_{12}$ (Fig.~\ref{fig:ind}C). However, the orientations of 
the existing segments of the trajectory do not determine the direction $\eta_x$, and $\ell_x$ can therefore 
be viewed as a ``provisional" alignment that may be actualized once the explicit information specifying its  
orientation emerges. Nevertheless, such alignments and the incidences that they induce may be incorporated 
into the hippocampal map, to accelerate its topological dynamics. 

As the turn continues, the next segment connects to $\sigma_4$ along $l_4=\{\sigma_3,\sigma_4|\eta_4\}$, 
allowing home run preplays $\ell_{40}$, $\ell_{41}$, $\ell_{42}$, which produce additional intersections, 
augmenting the lines $\ell_{30}$, $\ell_{31}$ and $\ell_{43}$ in specific order (Fig.~\ref{fig:ind}D).
Subsequent segments of the trajectory can generate ever larger sets of locations and alignments but the map
learning process can be terminated when the map $\mathscr{A}_{\mathcal{C}}(t_n)$ stabilizes topologically (see below).
At each step, the acquired collection of alignments embedded into the unfolding cognitive map sustains its
ongoing geometric structure.

Other alignments may be produced by more complex relationships within the existing configurations and their 
maps, as suggested, e.g., by Desargues or Pappus theorems. However, in finite planes these relationships may 
not be necessitated by the incidence axioms and require additional properties and implementing mechanisms
\cite{Hilbert,Vossen,Batten,Karteszi}. 
This scope of questions falls beyond this discussion and will be addressed elsewhere.
 
\textbf{Topological quantification of geometric learning}. The assumption of the model is that the influx of 
endogenously generated $\sigma$- and $\zeta$-simplexes accelerates the emergence of an aligned cognitive map 
with the correct topological shape. Testing this hypothesis requires computing the persistent homologies of 
the corresponding \textit{aligned complexes} $\mathcal{T}_{\sigma}^{\ell}(t)$ and $\mathcal{T}_{\zeta}^{\ell}(t)$; 
however, the algorithm described above produces the locations $\sigma_i$ and their appearance times $t_i$ without 
specifying cells that comprise a given simplex or detailing how these cells are shared between simplexes,
which is required for the homological computations. 

To extract the needed information, consider a graph $\mathcal{G}_{\sigma}$ whose links correspond to the 
adjacent simplexes, i.e., vertexes $v_i,v_j\in \mathcal{G}_{\sigma}$ are connected if $\sigma_i\multimapdotboth
\sigma_j$. If each $\sigma_i$ acts as an assembly, i.e., ignites when all of its vertex-cells (\ref{sigma}) 
activate and if the adjacent simplexes share vertexes, i.e., $\sigma_i\multimapdotboth\sigma_j\Leftrightarrow
\sigma_i\cap\sigma_j\neq\varnothing$ (required for spatiotemporal contiguity, see \cite{SchemaM}), then each 
$\mathcal{G}_{\sigma}$-link marks at least one putative cell $c_k$ shared by $\sigma_i$ and $\sigma_j$. 
In a conservative estimate (assuming, e.g., no ``redundant" cells that manifest themselves within just one
assembly), the set of $\mathcal{G}_{\sigma}$-links terminating at a given vertex $\sigma$ thus defines the
neuronal decomposition (\ref{sigma}) of the corresponding simplex. Same analyses allow restoring neuronal
decompositions for $\eta$-simplexes and constructing the $\zeta$-simplexes, thus producing cliques simplicial
complexes $\mathcal{T}_{\sigma}^{\ell}(t)$, $\mathcal{T}_{\eta}^{\ell}(t)$ and $\mathcal{T}_{\zeta}^{\ell}(t)$.

If, according to the requirement \textbf{R1}, the resulting $\sigma$- and $\eta$-fields cover their respective
representing spaces $\mathcal{E}$ and $S^1$, then the $\zeta$-fields cover the orientation space $\mathcal{O}=
\mathcal{E}\times S^1$, and the nerves associated with these covers, along with their temporal representations,
$\mathcal{T}_{\sigma}^{\ell}(t)$ and $\mathcal{T}_{\zeta}^{\ell}(t)$, should have the required topological 
properties. However, this argument has a principal caveat: some locations induced through endogenous network
activity may correspond to physically inaccessible domains in $\mathcal{E}$, which may divert the evolution of
the resulting coactivity complex from the topology of the place field nerve of the navigated environment.
Simulations show that indeed, the ``autonomously constructed" complex $\mathcal{T}_{\sigma}^{\ell}(t)$ tends 
to acquire a trivial shape ($b_{n>0}(\mathcal{T}_{\sigma}^{\ell})=0$) irrespective of the shape of the 
underlying $\mathcal{E}$.

A solution to this problem may be based on exploring functional differences between place cell combinations
$\dot{\sigma}_i$ that represent ``physically allowed" locations and the combinations $\acute{\sigma}_k$ that
represent ``physically prohibited" regions. One would expect that in a confined environment, the former kind
of cell groups should reactivate regularly due to animal's (re)visits, whereas the latter kind is never 
``validated" through actual exploration---$\acute{\sigma}_k$s may activate only during the occasional preplays
or replays. Taking advantage of this difference, let us assume that cell assemblies have a finite lifetimes
\cite{Syntax,Harris}, i.e., that 1) the probability of an assembly's disappearance after an inactivity period
$t_{\sigma}$ is $$p_{\sigma}\simeq e^{-t_{\sigma}/\tau_{\sigma}},$$ where $\tau_{\sigma}$ is $\sigma$'s mean
decay period, and 2) that the decay process resets ($t_{\sigma}=0$) after each reactivation of $\sigma$ (for 
some physiological motivations and references see \cite{PLoZ,Replays}).

To emphasize the contribution of the locations imprinted into the network structure due to physical activity 
over the computationally induced locations, the latter may be attributed with a shorter decay period, $\tau_{
	\acute{\sigma}}=\tau\ll T_{\min}^{\sigma}$, whereas the former may be treated as semi-stable $\tau_{\dot{
	\sigma}}\gg\tau_{\acute{\sigma}}$, e.g., for basic estimates, one can use $\tau_{\dot{\sigma}}=\infty$.
Lastly, the transition between $\acute{\sigma}$s and $\dot{\sigma}$s is modeled by stabilizing the decaying 
assemblies upon validation, i.e.,
\begin{equation}
\tau_{\sigma} = \begin{cases}
\tau_{\acute{\sigma}}=\tau & \mbox{before animal visits } \upsilon_{\sigma}, \\ 
\tau_{\dot{\sigma}}=\infty & \mbox{after animal visits } \upsilon_{\sigma}. 
\end{cases}
\nonumber
\end{equation} 
With this plasticity rule, physically permitted locations $\dot{\sigma}$ should maintain their presence in
the map, whereas the prohibited locations $\acute{\sigma}$ should decay, revealing the physical shape of the 
environment (Fig.~\ref{fig:ind}E).

To verify this approach, the semi-random foraging trajectory simulated in Section~\ref{sec:topol} was replaced
with a polygonal chain trajectory consisting of straight moves and random yaws. The preplays were then modeled
by injecting straight alignments into the coactivity graph as soon as the required information became available
(for details see \cite{Replays}). Based on the results of \cite{PLoZ,Replays}, the decay rate $\tau=0.5$ secs 
was selected to model the dynamics of the unstable locations.

For these parameters, the homological characteristics of the resulting ``flickering" coactivity complex
evaluated using ZigZag Homology techniques \cite{Carlsson2,Carlsson3,EdelsbrunnerZ}, quickly became stable: the
Betti numbers stabilized at $b_{n\leq1}(\mathcal{T}_{\zeta}^{\ell})=1$, $b_{n>1}(\mathcal{T}_{\zeta}^{\ell})=0$,
in $T_{\min}^{\zeta}\approx 5$ minutes, which approximately matches the hippocampal learning time 
$T_{\min}^{\sigma}$ and demonstrates that geometric organization of the cognitive map brings learning dynamics
to the biologically viable timescale.

\section{Discussion}
\label{sec:disc}

The proposed models of orientation learning are built by combining inputs from the hippocampal place cells and
the head directions cells. Experiments demonstrate that these two populations of neurons are coupled: in slowly 
deforming environments, their spiking activities remain highly correlated, pointing at a unified cognitive spatial
framework that involves both locations and orientations \cite{Knierim,Yoganarasimha1,Hargreaves,Sargolini}. 
The goal of this study is to combine topological and geometric approaches to model such a framework, and to 
evaluate the corresponding learning dynamics. 

The first model (Section~\ref{sec:topol}) is based on the observation that both the hippocampal and the head
direction maps are of a topological nature: while the place cells encode a qualitative, elastic map of the
navigated environment $\mathcal{E}$ \cite{Gothard1,Alvernhe1,Alvernhe2,Alvernhe3,eLife,Wu}, the head direction
cells map the space of directions, $S^1$ \cite{Taube2}. A combination of place and head direction cells' inputs
can hence be used to construct an extended topological map of oriented locations $\mathcal{O}$, which has a
structure of a direct product $\mathcal{E}\times S^1$---a natural framework for describing the kinematics of
rats' movements.

The second model (Section~\ref{sec:geom}) is structurally similar (a discrete map of directions is associated 
with each location), but involves constructions that define an additional, geometric layer of the cognitive 
map's architecture. In particular, this model allows viewing spatial orientation learning from a geometric 
perspective---not only as a process of discovering connections between locations, but also establishing shapes
of location arrangements, e.g., straight or turning paths, their incidences, intersections, junctions, etc. 
In neuroscience literature, such references are commonly made in relation to the physical geometry of the 
representing spaces $\mathcal{E}$ and $S^1$, e.g., the ``straightness" of a $\sigma$-field arrangement implies
simply that it can be matched by a Euclidean line in the environment where the rat is observed \cite{Valerio,McNPth}.
However, understanding the geometric structure of the cognitive map requires interpreting neuronal activity
in systems' own terms, rather than through the parameters of exterior geometry.

The key observation underlying the geometric model is that the activity of head direction cells ``tags'' the 
activity of place cells in a way that allows an \textit{intrinsic} geometric interpretation of the combined 
spiking patterns, i.e., defining alignments, turns, yaws, etc., in terms of neuronal spiking parameters. A 
famous quote attributed to D. Hilbert proclaims that ``\textit{the axioms of geometry would be just as valid
	if one replaced the undefined terms `point, line, and plane' with `table, chair, and beer mug'...}" 
\cite{Blumenthal}. 
From such perspective, this model aims at constructing a synthetic ``location and compass" neuro-geometry in
terms of the temporal relationships between the spike trains, without using extrinsic references or \textit{ad
	hoc} measures, which may be a general principle for how space and geometry emerge from neuronal activity.

In order to emphasize connections with conventional geometries, the model is formulated in a semi-axiomatic
form. However, in contrast with the standard affine \textbf{A}-axioms or their direct analogues, the 
\textbf{A}$_n$-axioms, the rules \textbf{G1}--\textbf{G3} serve not just as formal assertions that lay logical
foundations for geometric deductions, but also as reflections of physiological properties of the networks that 
implement the computations. First, since the networks contain a finite number of neurons and may actualize a
finite set of locations and alignments, the emergent geometry is finitary. Second, certain notions that in 
standard discrete affine plane $\mathscr{A}$ are introduced indirectly, \textit{relationally}, become 
\textit{constructive} in the ``cognitive" affine plane. For example, directions defined through equivalence
classes of parallel lines in $\mathscr{A}$ \cite{Hilbert,Vossen,Batten,Karteszi}, are defined explicitly in 
$\mathscr{A}_{\mathcal{C}}$ using the directing $\eta$-activities. Third, certain elements of the geometric
structure are actualized explicitly through the network's architecture, e.g., a fixed number of alignments 
passing through every location is implemented by cell assemblies wired into the head direction network 
\cite{Bassett,Reddish1}. Other properties are not prewired but \textit{acquired} during a particular learning
experience and reflect both the physical structure of a specific environment and intrinsic mechanisms of spatial 
information processing.

\section{Methods}
\label{section:methods}

The computational algorithms used in this study were described in \cite{PLoS,Arai,Basso,Hoffman,SchemaS,SchemaM,CAs,PLoZ,Replays}.

\textbf{The environment} shown on Fig.~\ref{fig:PFmap}A is similar to the arenas used in electrophysiological
experiments \cite{Hafting,BrunG}. The simulated trajectory represents exploratory spatial behavior that does 
not favor one segment of the environment over another.

\textbf{Place cell spiking} probability was modeled as a Poisson process with the rate 
\begin{equation*}
	\lambda(r)=f e^{-\frac{(r-r_0)^2}{2s^2}},
	\nonumber
\end{equation*}
where $f$ is the maximal rate and $s$ defines the size of the firing field centered at $r_0 = (x_0, y_0)$ 
\cite{Barbieri}. In addition, spiking probability was modulated by the $\theta$-waves, which also define 
the temporal window $w \approx 250$ ms (about two $\theta$-periods) for detecting the place cell spiking
coactivity \cite{Arai,Mizuseki}. The place field centers $r_0$ for each computed place field map were 
randomly and uniformly scattered over the environment.

\textbf{Persistent Homology Theory} computations were performed using \textsf{Javaplex} computational software
\cite{javaplex} as described in \cite{PLoS,Arai,Basso,Hoffman,CAs}. Usage of Zigzag persistent homology 
methods is described in \cite{PLoZ,Replays}.

\vspace{7mm}
\textbf{Acknowledgments}. The author is grateful to D. Morozov for providing Zigzag persistent homology 
simulating software. The work was supported by the NSF grant 1901338.

\textbf{COI statement}: The author states that there is no conflict of interest.


\end{document}